# Efficiency of Matrix Multiplication on the Cross-Wired Mesh Array

Subhash Kak

**ABSTRACT**

This note looks at the efficiency of the cross-wired mesh array in the context of matrix multiplication. It is shown that in case of repeated operations, the average number of steps to multiply sets of $n \times n$ matrices on a mesh array of size $n^2$ approaches $n$.

## INTRODUCTION

There is continuing interest in speeding up matrix multiplication which is one of the most basic computing operations. There is also interest in the multiplication of matrices with specific properties and for particular applications. (e.g. [1]-[4]). Recently, Bae et al. [5] provided a summary of the efforts in this area together with their own algorithm on the 2D mesh array of size $n^2$ that computes the product C = AB of two $n \times n$ matrices in $1.5n – 1$ steps. These results are valid for the standard mesh array of Figure 1, which presents an example of matrix multiplication [6]. The standard array requires $(3n-2)$ steps to compute the product of two $n \times n$ matrices. The more efficient result is obtained if the data is so loaded on the array that no computing cells are idle. The data is loaded in all four directions of the 2D array and some additional registers are required in each computing cell to manage this data.

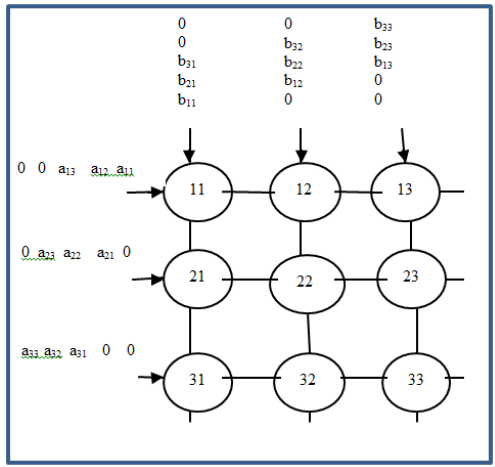

Figure 1. The standard 2-D mesh array for matrix multiplication

In contrast to the standard 2D mesh array of Figure 1 stands the *cross-wired mesh array* of Figure 2 that we call 2DC (cross-wired 2D). The structure of the computing cell in 2DC is shown in Figure 3. Each cell receives data from NW (northwest) and NE (northeast) registers and after the data is multiplied and added to the accumulator in the middle called Prod Sum (product sum), it is passed to SE (southeast) and SW (southwest) registers, respectively. After the central register has completed its accumulation, the result is transported out by the horizontal links – either to the left or the right -- and the cell can be reused for another computation.

The 2DC array is slightly different from the standard 2D array. The cross-wired mesh architecture [7][8] (Figure 1) was proposed to speed up the computation for multiplying two $n \times n$ matrices. In Figure 2, we have the cross-wired mesh array for multiplying two 4×4 matrices which takes 7 steps, whereas the standard systolic array [1] requires the same number of steps to multiply two 3×3 matrices. In the general case of $n\times n$ matrices, the cross-wired mesh requires *(2n-1)* steps which are a substantial speed-up over the performance of the standard array. The speedup of the mesh array is a consequence of the fact that no zeros are padded in its inputs. The larger question of whether relatives of 2D arrays [9] provide fundamental advantage in implementation of certain algorithms may be asked. Clearly as shown earlier the 2DC array is more efficient than the 2D array, but we don't know if this advantage carries forth to algorithms other than that of matrix multiplication.

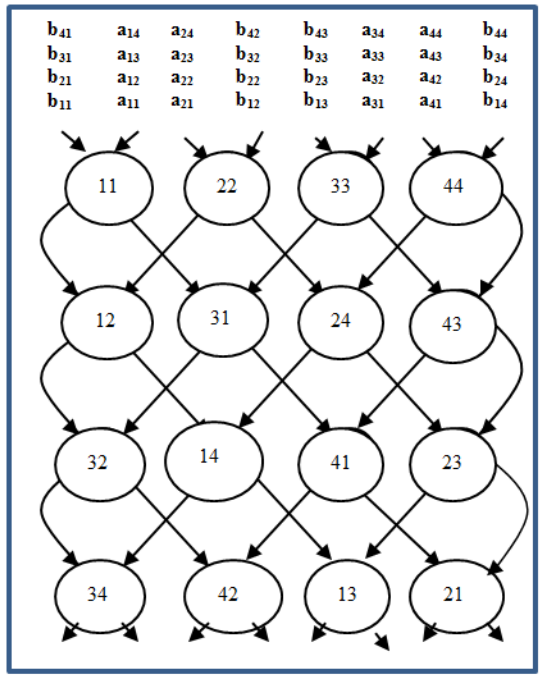

Figure 2. The 2D cross-wired mesh architecture for multiplying two 4×4 matrices; the numbers within each cell compute components of the product matrix. Thus 14 means $c_{14}$.

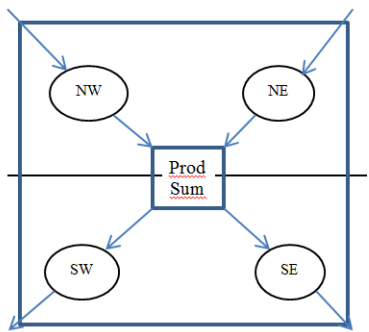

Figure 3. The structure of a computing cell in the cross-wired mesh array

In this note we consider the performance of the cross-wired mesh array when a batch of matrix multiplications is to be performed. One motivation for this is the many applications where repeated matrix multiplications are used (e.g. [10],[11]). We wish to estimate of the efficiency of the 2DC mesh in the case of continuous operation.

# REPEATED OPERATION OF 2DC MESH ARRAY

Let us consider repeated matrix operations on the *4×4* array of Figure 2. We begin by loading the top row of the array at time t=0 and subsequently update the contents of the central accumulation register at further time steps. The contents of the 16 cells in the two arrays will update as follows:

$$t=0: \begin{bmatrix} a_{11}b_{11} & a_{21}b_{12} & a_{31}b_{13} & a_{41}b_{14} \\ 0 & 0 & 0 & 0 \\ 0 & 0 & 0 & 0 \\ 0 & 0 & 0 & 0 \end{bmatrix};$$

$$t=1: \begin{bmatrix} a_{11}b_{11}+a_{12}b_{21} & a_{21}b_{12}+a_{22}b_{22} & a_{31}b_{13}+a_{32}b_{23} & a_{41}b_{14}+a_{42}b_{24} \\ a_{11}b_{12} & a_{31}b_{11} & a_{21}b_{14} & a_{41}b_{13} \\ 0 & 0 & 0 & 0 \\ 0 & 0 & 0 & 0 \end{bmatrix};$$

$$t=2: \begin{bmatrix} a_{11}b_{11}+a_{12}b_{21}+a_{13}b_{31} & a_{21}b_{12}+a_{22}b_{22}+a_{23}b_{32} & a_{31}b_{13}+a_{32}b_{23}+a_{33}b_{33} & a_{41}b_{14}+a_{42}b_{24}+a_{43}b_{34} \\ a_{11}b_{12}+a_{12}b_{22} & a_{31}b_{11}+a_{32}b_{21} & a_{21}b_{14}+a_{22}b_{24} & a_{41}b_{13}+a_{42}b_{23} \\ a_{31}b_{12} & a_{11}b_{14} & a_{41}b_{11} & a_{21}b_{13} \\ 0 & 0 & 0 & 0 \end{bmatrix};$$

t=3:
$$\begin{bmatrix} a_{11}b_{11}+a_{12}b_{21}+a_{13}b_{31}+a_{14}b_{41} & a_{21}b_{12}+a_{22}b_{22}+a_{23}b_{32}+a_{24}b_{42} & a_{31}b_{13}+a_{32}b_{23}+a_{33}b_{33}+a_{34}b_{43} & a_{41}b_{14}+a_{42}b_{24}+a_{43}b_{34}+a_{44}b_{44} \\ a_{11}b_{12}+a_{12}b_{22}+a_{13}b_{32} & a_{31}b_{11}+a_{32}b_{21}+a_{33}b_{31} & a_{21}b_{14}+a_{22}b_{24}+a_{23}b_{34} & a_{41}b_{13}+a_{42}b_{23}+a_{43}b_{33} \\ a_{31}b_{12}+a_{32}b_{22} & a_{11}b_{14}+a_{12}b_{24} & a_{41}b_{11}+a_{42}b_{21} & a_{21}b_{13}+a_{22}b_{23} \\ a_{31}b_{14} & a_{41}b_{12} & a_{11}b_{13} & a_{21}b_{11} \end{bmatrix}$$

At t=4, the top row Prod Sum register contents will be moved using the horizontal line out of the array, and the components of the two new matrices will begin to be loaded into the array row by row. For the product C=AB, the components will arrive in the following order:

```
11  22  33  44
12  31  24  43
32  14  41  23
34  42  13  21
```

The order of arrival of the components by taking it out from the left when written in the standard form is as follows:

$$\begin{bmatrix} 1 & 5 & 15 & 10 \\ 16 & 2 & 12 & 7 \\ 6 & 9 & 3 & 13 \\ 11 & 14 & 8 & 4 \end{bmatrix}$$

We see that the fourth row is the mirror reversed image of the second row. Also, the third row has symmetry within itself. In general for *n* even, the rows *2* to *n/2* are mirror image to rows *n/2+2* to *n*, and the middle row *(n/2 +1)* has self-symmetry.

As shown in [2], given the matrix of values for *n×n*, the array for the *(n+1)×(n+1)* can be developed by inspection exploiting the various symmetries in the structure. This may be seen in the transition from the 6×6 case to the 7×7 case:

```
11 22 33 44 55 66 77
12 31 24 53 46 75 76
32 14 51 26 73 47 65
34 52 16 71 27 63 45
54 36 72 17 61 25 43
56 74 37 62 15 41 23
76 57 64 35 42 13 21
```

The bold values are the only ones that are new as the remainder is determined by the internal symmetries mentioned earlier. The bold values themselves are determined by the symmetries along the several diagonals as the second subscript there is fixed in a specific diagonal just as the first subscript is fixed along alternate anti-diagonal.

## EFFICIENCY OF 2DC MESH ARRAY

Let us assume that there are N pairs of *n×n* matrices that need to be multiplied. Observe the following:

1. It takes for each matrix pair a total of *(2n-1)* steps to load and exit the array.
2. If the computation began at t=0, the N pairs of matrices would all have been computed by *t=Nn+n-1* steps.
3. Certain number of cells remains idle during initial loading and the eventual completion of the computation.

The number of unused cells in the initial loading is the sum of the inactive rows times the number of cells in each row:

$$\frac{n(n-1)n}{2}$$

The number of inactive cells at the end of the computation equals the sum of the rows ranging from 1 to n-1 times the number of elements in each row:

$$\frac{n(n-1)n}{2}$$

The total number of inactive cells in the computation process is $n^3 - n^2$. During this period the total number of cells to be accounted for is

$$(N+1)n^3 - n^2$$

Therefore, the number of cells that are in active use during the computation process is:

$$(N+1)n^3 - n^2 - n^3 + n^2 = Nn^3$$

The efficiency of the array is therefore:

$$\frac{Nn^3}{(N+1)n^3 - n^2} = \frac{N}{N+1-n^{-1}}$$

As the number of operations, *N*, increases, the efficiency goes to 1.

## DISCUSSION

This note shows that although the computed product matrix values will still take *(2n-1)* steps for any pair of matrices, the average number of steps approaches *n* as the number of repeated operations increases. The cross-wired mesh array therefore provides better results than the standard mesh array.